\documentclass[aps,prl,groupedaddress]{revtex4}
\usepackage{graphicx}

\def\be{\begin{equation}}
\def\ee{\end{equation}}
\def\ba{\begin{eqnarray}}
\def\ea{\end{eqnarray}}
\def\bc{\begin{center}}
\def\ec{\end{center}}

\begin{document}

\title{A new electromagnetic mode in graphene}

\author{S. A. Mikhailov}
\email[Electronic mail: ]{sergey.mikhailov@physik.uni-augsburg.de}
\affiliation{Institut f\"ur Physik, Universit\"at Augsburg, D-86135 Augsburg, Germany}

\author{K. Ziegler}
\affiliation{Institut f\"ur Physik, Universit\"at Augsburg, D-86135 Augsburg, Germany}

\date{\today}

\begin{abstract}
A new, weakly damped, {\em transverse} electromagnetic mode is predicted in graphene.
The mode frequency $\omega$ lies in the window $1.667<\hbar\omega/\mu<2$, where $\mu$ is the chemical potential, 
and can be tuned from radiowaves to the infrared by changing the density of charge carriers through a gate voltage. 
\end{abstract}


\maketitle

In the past two years, a great deal of attention has been attracted by the discovery of graphene,
a truly two-dimensional (2D) electronic system \cite{Novoselov05,Zhang05} (for recent 
reviews see Refs. \cite{Katsnelson07,Geim07a}). Graphene is a monolayer of carbon atoms, and the 
band structure of electrons in graphene consists of six Dirac cones at the corners of the hexagon-shaped
Brillouin zone \cite{Wallace47,Semenoff84}, with the massless, linear electron/hole dispersion with 
the effective ``velocity of light'' $V\approx 10^8$ cm/s. The special spectrum of the charge carriers leads to 
a number of interesting transport properties, which have been intensively studied in the literature, see e.g. Refs.
\cite{Gusynin05,Ziegler06,Katsnelson06,Cheianov06,Gusynin06a,Falkovsky06,Nilsson06,Nomura07,Fuchs07,Silvestrov07,
Novoselov07} and review articles \cite{Katsnelson07,Geim07a}. 

Electrodynamic properties of graphene have been studied in Refs. 
\cite{Gusynin06a,Gusynin06b,Falkovsky06,Abergel06,Hanson07,Apalkov06,Hwang06,Vafek06,Wunsch06,Iyengar06,Trauzettel07}. 
Frequency-dependent conductivity \cite{Gusynin06a,Gusynin06b,Falkovsky06,Abergel06,Hanson07} and collective excitations 
of the graphene layer -- plasmons \cite{Apalkov06,Hwang06}, thermoplasma polaritons \cite{Vafek06} -- 
have been theoretically investigated. It has been shown that the specific band structure of graphene leads to 
a certain modification of the plasmon and plasmon-polariton spectra, as compared to conventional 2D electron systems
with the parabolic dispersion of electrons.

In this Letter we show that the Dirac spectrum of electrons leads to a radically new feature of the electrodynamic 
response of the electron-hole plasma in graphene, as compared to conventional electron systems. We predict 
the existence of a {\em transverse} (TE) electromagnetic mode in graphene, the mode which cannot exist in 
systems with the parabolic electron dispersion. The new mode propagates along the graphene layer with the 
velocity close to the velocity of light, has a weak damping, and its frequency is tunable across a broad frequency 
range from radiowaves to the infrared. These properties may have a strong potential for future electronic and 
opto-electronic applications of graphene. 

To explain the essense of our finding, we briefly discuss the nature of electromagnetic modes in conventional 
electronic systems. Isotropic and uniform three-dimensional plasmas can support (in zero magnetic field) both 
longitudinal and transverse electromagnetic modes. The electric field vector in the longitudinal (transverse)  
wave is parallel (perpendicular) to the wave vector. In the 2D electron gas, however, only the longitudinal 
(or transverse magnetic, TM) modes -- 2D plasmons and plasmon-polaritons -- may exist under standard experimental
conditions \cite{noteChaplik72}. The spectrum of electromagnetic modes, propagating along and localized near 
the 2D electron gas layer, has the form
\be
1+\frac{2\pi i \sigma(\omega)\sqrt{q^2-\omega^2/c^2}}{\omega}=0
\label{TM}
\ee
for the TM waves \cite{Stern67}, and 
\be
1-\frac{2\pi i \omega \sigma(\omega)}{c^2\sqrt{q^2-\omega^2/c^2}}=0
\label{TE}
\ee
for the transverse electric (TE) waves \cite{Falko89}; here $\sigma(\omega)$ is the local dynamic conductivity
of the 2D gas and $c$ is the velocity of light. As seen from (\ref{TM}) and (\ref{TE}), the TM (TE) modes may exist
if and only if the imaginary part of $\sigma(\omega)$ is positive (negative). In conventional 2D electron systems,
realized, for instance, in GaAs/AlGaAs quantum-well structures, the conductivity can be described by the Drude model
$\sigma(\omega)=in_se^2/m(\omega+i\gamma)$, where $n_s$, $e$, $m$ and $\gamma$ are the density, the electric charge,
the effective mass and the scattering rate of 2D electrons, respectively. As $\sigma''(\omega)>0$, only the TM waves
(plasmons, plasmon-polaritons) can propagate in such structures.

\begin{figure}
\includegraphics[width=8.5cm]{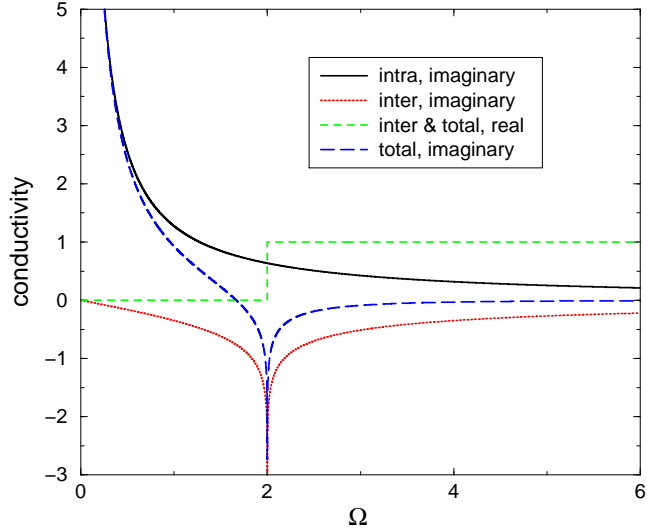}
\caption{\label{condg0T0} (Color online) The dynamic conductivity of the graphene layer, in units $e^2g_sg_v/16\hbar$, 
as a function of the frequency $\Omega=\hbar\omega/\mu$ in the collisionless limit 
at zero temperature $T/\mu=0$. Imaginary and real parts of the intra-band and inter-band contributions, as well 
as their sum (the total conductivity) are shown.
}
\end{figure}

In graphene with the massless Dirac form of the electron/hole dispersion the situations is different. To demonstrate this,
we begin with the Hamiltonian $\hat H=V\sigma_\alpha\hat p_\alpha$, which determines the energy spectrum of charge
carriers in the electron ($l=2$) and hole ($l=1$) bands, 
\be
E_{{\bf k}l}=(-1)^l \hbar Vk, \ \ l=1,2,
\ee
and the corresponding wave functions $|{\bf k}l\rangle$. 
Here $\alpha=(x,y)$, $\sigma_\alpha$ are Pauli matrixes, $\hat p_\alpha$ is the momentum operator, and 
${\bf k}=(k_x,k_y)$, $k=\sqrt{k_x^2+k_y^2}$ is the vave vector. Using the self-consistent-field approach 
\cite{Ehrenreich59} (alternatively, the Kubo formalism \cite{Ziegler06}, or the random-phase approximation \cite{Platzman73}) 
we calculate the high-frequency conductivity of the system 
$\sigma_{\alpha\beta}(\omega)=\sigma(\omega)\delta_{\alpha\beta}$. Scattering is ignored (i.e. $\gamma=0$) here. 
The conductivity consists of the intra-band and the inter-band contributions \cite{Gusynin06b,Falkovsky06}, 
\be
\sigma_{\alpha\beta}^{intra}(\omega)
=
\frac {-ie^2}{\hbar^2(\omega+i0) S}\sum_{{\bf k}l}
\frac{\partial E_{{\bf k}l}}{\partial k_\alpha}
\frac{\partial f(E_{{\bf k}l})}{\partial E_{{\bf k}l}}
\frac{\partial E_{{\bf k}l}}{\partial k_\beta},
\label{intra}
\ee
\be
\sigma^{inter}_{\alpha\beta}(\omega)
=
\frac {ie^2\hbar }{S}\sum_{{\bf k},l\neq l'}\frac
{f(E_{{\bf k}l'})-f(E_{{\bf k}l})}
{E_{{\bf k}l'}-E_{{\bf k}l}-\hbar(\omega+i0)}
\frac 1
{E_{{\bf k}l'}-E_{{\bf k}l}}
\langle {\bf k}l|\hat v_\alpha |{\bf k}l'\rangle 
\langle {\bf k}l'| \hat v_\beta  |{\bf k}l\rangle,
\label{inter}
\ee
where $f(E_{{\bf k}l})$ is the Fermi distribution function, $\hat v_\alpha=V\sigma_\alpha$ is the velocity operator,
and $S$ is the sample area. In the gate controlled graphene systems, where the density of charge carriers can be tuned
by the gate voltage, we obtain at $T/\mu\to 0$, that the intra-band conductivity (\ref{intra}) assumes the Drude-like form, 
\be
\sigma_{intra}(\omega)=\frac {in_se^2V^2 }{ (\omega+i0) \mu}=\frac{e^2g_sg_v}{16\hbar}\frac {4i}{\pi\Omega},
\label{intra1}
\ee
while the inter-band contribution (\ref{inter}) gives
\be
\sigma_{inter}(\omega)=\frac{e^2g_sg_v}{16\hbar}
\left(\theta \left(|\Omega|-2\right)-
\frac i\pi
\ln
\left|
\frac{2+\Omega}{2-\Omega}
\right|\right),
\label{inter1}
\ee
see Figure \ref{condg0T0}. 
Here $\Omega=\hbar \omega/\mu$, $\mu$ is the chemical potential, $n_s=g_sg_v\mu^2/4\pi\hbar^2V^2$ is the density 
of electrons at $T=0$, and $g_s=2$ and $g_v=2$ are the spin and valley degeneracies in graphene (six Dirac cones 
give the valley degeneracy $g_v=2$ because only one third of each cone belongs to the first Brillouin zone 
\cite{Semenoff84}).

As seen from Eqs. (\ref{intra1}) and (\ref{inter1}), as well as from Figure \ref{condg0T0}, the imaginary part of 
$\sigma_{intra}(\omega)$ is positive, like in the standard 2D electron systems, so that the intra-band contribution 
alone could not provide the conditions of the TE mode existence. The imaginary part of the inter-band contribution, 
however, is negative and diverges logarithmically at $\Omega\to 2$. This divergency is associated with the 
step-like behavior of the real part $\sigma'_{inter}(\omega)$, which describes the inter-band absorption of 
radiation at $\hbar\omega>2\mu$. Using (\ref{intra1}) and (\ref{inter1}), the TE-mode dispersion (\ref{TE}) 
can now be written as 
\be
\sqrt{Q^2-\Omega^2}=\frac{g_sg_v}{4}\frac{e^2}{\hbar c}\left(\frac\Omega 2\ln\frac{2+\Omega}{2-\Omega}-2\right),\ \ \Omega<2, 
\label{TE1}
\ee
where $Q=\hbar c q/\mu$. Notice that apart from the dimensionless wave-vector $Q$ and the frequency $\Omega$, 
Eq. (\ref{TE1}) depends only on the fine structure constant $\alpha=e^2/\hbar c$. The TE mode (\ref{TE1}) 
doesn't decay at $T=0$ and exists in the window $1.667<\Omega<2$, where the term in brackets in (\ref{TE1}) 
is positive. To give a numerical example, at $n_s\simeq 6\times 10^{12}$ cm$^{-2}$ the TE mode should exist in 
the window 115 THz $\lesssim f\lesssim 140$ THz, at $n_s\simeq 10^{11}$ cm$^{-2}$ -- in the window 15 THz 
$\lesssim f\lesssim 18$ THz, and so on. Choosing an appropriate value of the gate voltage, one can tune the 
TE mode to a desired frequency range. This may be useful in designing devices for infrared, terahertz, 
and microwave electronics. The TM mode (2D plasmon-polariton) does not exist at $1.667<\Omega<2$; its 
dispersion relation (\ref{TM}) has real solutions only at $\Omega<1.667$. 

\begin{figure}
\includegraphics[width=7cm]{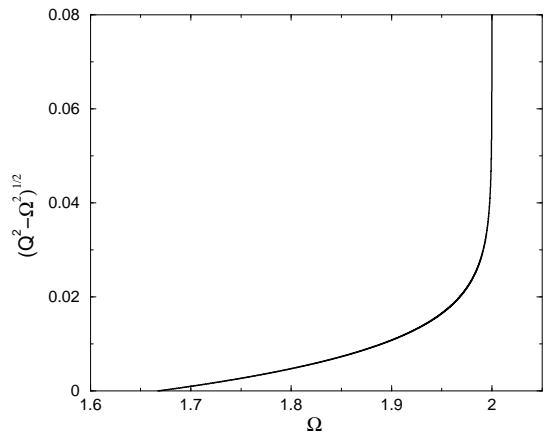}
\caption{\label{kappa} The value of $\sqrt{Q^2-\Omega^2}$ as a function of the dimensionless frequency $\Omega$ 
for the TE mode.}
\end{figure}

Figure \ref{kappa} illustrates the dependence of $\sqrt{Q^2-\Omega^2}$ on the frequency $\Omega$, Eq. (\ref{TE1}). 
Due to the small factor $\alpha=e^2/\hbar c$ in the right-hand side of (\ref{TE1}), the deviation of the wavevector 
$Q$ from the frequency $\Omega$ in the wave is small as compared to the values of $Q$ and $\Omega$ themselves. 
This means that the TE mode propagates along the 2D graphene layer ($x$ direction) with the velocity close to 
the velocity of light, $\omega\lesssim cq$, and that the localization length of the wave in the perpendicular 
($z$) direction is much larger than its wavelength in the $x$ direction.

\begin{figure}
\includegraphics[width=8.5cm]{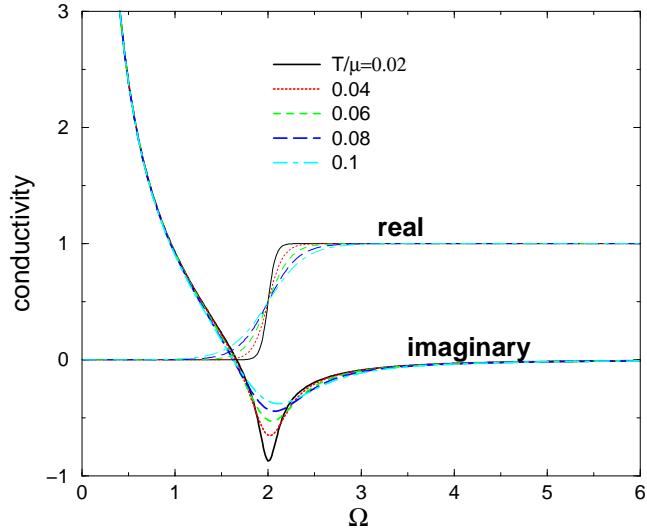}
\caption{\label{finT1} (Color online) The real and imaginary parts of the dynamic conductivity of graphene, in 
units $e^2g_sg_v/16\hbar$, as a function of frequency $\Omega=\hbar\omega/\mu$ at finite temperatures. 
}
\end{figure}

\begin{figure}
\includegraphics[width=8.5cm]{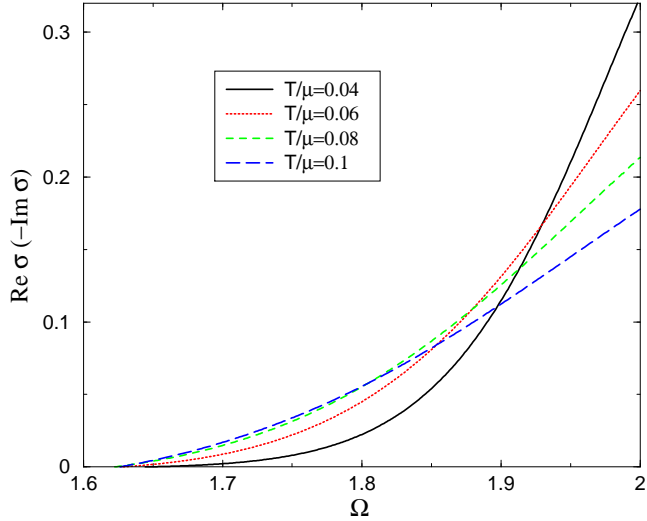}
\caption{\label{finT} (Color online) The factor $\tilde\sigma'(\omega)[-\tilde\sigma''(\omega)]$ from Eq. 
(\ref{Qfactor}) as a function of $\Omega=\hbar\omega/\mu$ at finite temperatures. 
}
\end{figure}

The frequency dependence of the conductivity at finite temperature is shown in Figure \ref{finT1}. At $T>0$, 
the real part $\sigma'(\omega)$ becomes finite at $\Omega<2$, and the TE mode acquires a finite damping, $Q=Q'+iQ''$. 
This damping, however, is very small. Using that $\alpha\ll 1$ we get from Eq. (\ref{TE1})
\be
\frac{Q''}{Q'}=\frac{q''}{q'}=\left(\frac{\pi\alpha}2\right)^2\tilde\sigma'(\omega)[-\tilde\sigma''(\omega)],
\label{Qfactor}
\ee
where $\tilde\sigma$ is the normalized conductivity of the layer, $\sigma=(e^2g_sg_v/16\hbar)\tilde\sigma$. 
Figure \ref{finT} exhibits the factor $\tilde\sigma'(\omega)[-\tilde\sigma''(\omega)]$ as a function of $\Omega$ 
at different temperatures. One sees that $\tilde\sigma'(\omega)[-\tilde\sigma''(\omega)]$ is smaller than 1 and 
hence that $q''/q'$ does not exceed $10^{-4}$ even at $T\simeq 0.1\mu$. At the density 
$n_s\simeq 6\times 10^{12}$ cm$^{-2}$ the Fermi energy $\mu$ corresponds to $T\simeq 3000$ K in graphene, 
so the TE mode should be easily observable at room temperature.

To conclude, we have predicted a new transverse electric mode in graphene.
The existence of the mode is directly related to the linear ``relativistic'' spectrum of charge carriers in graphene; 
in a conventional system of 2D electrons with the parabolic dispersion such a mode cannot exist. The mode frequency 
is widely tunable across the range from radiowaves to infrared frequencies, depending on the density of electrons or 
holes in the system. The damping of the mode is very weak even at room temperatures. 

The work was partly supported by the Swedish Research Council and INTAS.

\bibliography{../../BIB-FILES/mikhailov,../../BIB-FILES/lowD,../../BIB-FILES/dots,../../BIB-FILES/graphene,../../BIB-FILES/thz,../../BIB-FILES/zerores}

\end{document}